\documentstyle[preprint,prb,aps,epsf]{revtex}
\begin{document}
\draft
\title{Competing effects of mass anisotropy and spin Zeeman coupling on
the upper critical field 
of a mixed $d$- and $s$-wave superconductor}
\author{Wonkee Kim$^{1,2}$, Jian-Xin Zhu$^{1}$ and C. S. Ting$^{1,2}$}
\address{$^{1}$Department of Physics and Texas Center for Superconductivity,
University of Houston, Houston, Texas 77204\\
$^{2}$National Center for Theoretical Sciences, Hsinchu, Taiwan, 
Republic of China}
\maketitle
\begin{abstract}
Based on the linearized Eilenberger equations, 
the upper critical field $(H_{c2})$ of mixed $d$- and $s$-wave 
superconductors has been microscopically studied 
with an emphasis on the competing effects of mass anisotropy and  
spin Zeeman coupling. We find the mass anisotropy always 
enhance $H_{c2}$ while
the Zeeman interaction suppresses $H_{c2}$. 
As required by the thermodynamics, we find $H_{c2}$ is saturated 
at zero temperature.
We compare the theoretical calculations with
recent experimental data of YBa$_{2}$Cu$_{3}$O$_{7-\delta}$.
\end{abstract}
%PACS numbers: 74.20.-z,74.20.Mn,74.25.Bt

It has been a consensus~\cite{cons} that a high-$T_{c}$ 
cuprate superconductor has a $d$-wave
pairing symmetry, and CuO$_{2}$ plane is responsible for superconductivity. 
However, a pure $d$-wave
symmetry is appropriate only for a tetragonal lattice structure,
an orthorhombic material such as YBa$_{2}$Cu$_{3}$O$_{7-\delta}$ (YBCO)
is believed to have a subdominant $s$-wave component in the order parameter.
The orthorhombicity in YBCO is originated from a mass anisotropy (MA)
along $a$- and $b$-directions. In other words, $m_{a}$ is larger than $m_{b}$.
Such a discrepancy is mainly due 
to CuO chains in the $b$-direction.~\cite{carbotte}

Based on the Ginzburg-Landau theory, Xu {\it et al.}~\cite{jxu} 
have explored effects 
of the mass anisotropy to show that an $s$-wave component always coexists 
with a dominant $d$-wave component in the bulk order parameter. 
Belzig {\it et al.}~\cite{belzig} have obtained
the phase diagram of a mixed $d$- and $s$-wave superconductor
in the quasiclassical theory,~\cite{eilenberger} and
shown that the mass anisotropy gives rise to a non-zero $s$-wave component.

In this paper, we shall investigate effects of the MA
on the upper critical field $(H_{c2})$ of a mixed $d$- and $s$-wave 
superconductor, based on the quasiclassical theory,
using a quantum mechanical method
we have developed in Ref.~\cite{kim} 
and compare a theoretical result with recent experimental 
data.~\cite{nakagawa} We also take into account the paramagnetic Zeeman 
interaction (ZI) because $H_{c2}$ of a high-$T_{c}$ superconductor is large
at low temperature. 
As in Ref.~\cite{jxu,belzig}, we neglect
any other effects associated with the chains. For example,
we assume that lattice constants along $a$- and $b$-directions
have the same value.

Since the calculation
of $H_{c2}$ is a quantum mechanical problem of a charged particle in a 
constant magnetic field,~\cite{kim}
first of all we need to check if solvable is
a problem of a charged particle
with two different effective masses along $x$- and $y$-directions,
$m_{x}$ and $m_{y}$, respectively, in a constant magnetic field 
${\bf H}=\nabla\times {\bf A}=H\hat{\bf z}$.
In the symmetric gauge, the Hamiltonian ${\cal H}$ of the particle is 
given by
\begin{eqnarray}
{\cal H}&&={1\over 2m_{x}}(p_{x}+{e\over2}Hy)^{2}+
{1\over 2m_{y}}(p_{y}-{e\over2}Hx)^{2}
\nonumber\\
&&={1\over 2m_{x}}(p_{x}^{2}+\lambda p_{y}^{2})+{eH\over 2m_{x}}
(yp_{x}-\lambda xp_{y})+{e^{2}H^{2}\over 8m_{x}}(y^{2}+\lambda x^{2}),
\end{eqnarray}
where $\lambda=m_{x}/m_{y}$.
Introducing operators $a_{x}$ and $a_{y}$ such that
\begin{equation}
a_{x}={1\over\sqrt{2m_{x}\omega_{0}}}
(m_{x}\omega_{0}\sqrt{\lambda}x+ip_{x}),
\end{equation}
and
\begin{equation}
a_{y}={1\over\sqrt{2m_{x}\omega_{0}}}
(m_{x}\omega_{0}y+i\sqrt{\lambda}p_{y}),
\end{equation}
where $\omega_{0}=|e|H/2m_{x}$, ${\cal H}$ becomes
\begin{equation}
{\cal H}=\omega_{0}[a_{x}^{+}a_{x}+a_{y}^{+}a_{y}+\sqrt{\lambda}
+i(a_{x}^{+}a_{y}-a_{y}^{+}a_{x})].
\end{equation}
Let us introduce a new operator 
$b=(a_{x}+ia_{y})/\sqrt{2\sqrt{\lambda}}$
to simplify ${\cal H}$;
then, we obtain
\begin{equation}
{\cal H}=\omega_{c}\left( b^{+}b+{1\over2}\right),
\end{equation}
with $[b,b^{+}]=1$, $[b,{\cal H}]=\omega_{c}b$, and
$[b^{+},{\cal H}]=-\omega_{c}b^{+}$, 
where $\omega_{c}=|e|H/\sqrt{m_{x}m_{y}}$.
It shows that the problem we are considering reduces to that of a simple 
harmonic oscillator, and consequently $H_{c2}$ for an
anisotropic mixed $d$-and $s$-wave superconductor can be exactly calculated.

In the calculation of $H_{c2}$, we choose $[110]$ and $[-110]$ as $x$ and $y$
axis, respectively. An electron spectrum $\epsilon({\bf k})$ then changes
from $k_{a}^{2}/2m_{a}+k_{b}^{2}/2m_{b}$ to $k^{2}/2m+ 
ck_{x}k_{y}/m$. Here we have defined 
$m_{a}=m/(1-c)$ and $m_{b}=m/(1+c)$, where a small quantity $c$ is 
introduced to represent the degree of mass anisotropy. The Fermi surface 
(FS) now is elliptical, and the density of state $N(\phi)$ on FS becomes
$N_{0}(0)/[1+c\sin(2\phi)]$. 
By considering the anisotropy effect as a perturbation, we may assume that 
the order parameter can be still
expanded in terms of the set of unperturbed eigenstates $\{f_{N}({\bf R})\}$.
In Ref.~\cite{kim}, we have shown that the (singlet) 
order parameter is written 
as $A^{(d)}_{0}|0\rangle_{d}+A^{(s)}_{2}|2\rangle_{s}
+A^{(d)}_{4}|4\rangle_{d}$ near $H_{c2}$, where we denote $|N\rangle_{d(s)}$
as eigenstates $f_{N}$ of $d(s)$ channel. 
However, we expect that
some other states such as $|0\rangle_{s}$, $|1\rangle_{s(d)}$, and
$|2\rangle_{d}$ may involve in the order parameter due to the perturbation.
The linear combination of these states for the order parameter near $H_{c2}$
will be determined
by the symmetry of the system.

The linearized Eilenberger equation is written as:~\cite{rieck}
\begin{equation}
{\hat L}f(\omega,{\bf k},{\bf R})
+i\mbox{sgn}(\omega)\mu^{*}_{0}{\bf H}\cdot\bigl[
{\boldmath{\mbox{$\sigma$}}}f(\omega,{\bf k},{\bf R})
-f(\omega,{\bf k},{\bf R})
{\boldmath{\mbox{$\sigma$}}}^{tr}\bigr]
=2\pi\Delta({\bf R},{\bf k})\;,
\end{equation}
with
\begin{equation}
\Delta({\bf R},{\bf k})=T\sum_{\omega}\langle
V({\bf k},{\bf k}')f(\omega,{\bf k}',{\bf R})
\rangle_{FS}\;,
\label{gap}
\end{equation}
and
\begin{equation}
{\hat L}=2|\omega|+i\mbox{sgn}(\omega) {\bf v}_{F}\cdot{\bf \Pi}.
\end{equation}
Here $f(\omega,{\bf k},{\bf R})$ is a quasiclassical Green's function,
${\boldmath{\mbox{$\sigma$}}}=(\sigma_{x},\sigma_{y},\sigma_{x})$ are
Pauli matrices,  $\omega$ is the 
Matsubara frequency, $\mu^{*}_{0}$ can be interpreted as an effective 
magnetic moment of a quasi-electron with a mass anisotropy $(m_{a}\ne 
m_{b})$, which will be
considered as a phenomenological parameter associated with a coupling strength
between an electron spin and a magnetic field. The symbol $\langle \cdots 
\rangle_{SF}=\int {d\phi^{\prime}\over 2 \pi} N(\phi^{\prime}) 
\cdots$ represents the angular average over the Fermi surface.     
Since ${\bf v}_{F}=v_{Fa}\hat{\bf a}+v_{Fb}\hat{\bf b}$ and 
${\bf \Pi}=-i\nabla_{\bf R}-2e{\bf A}=\Pi_{a}\hat{\bf a}+\Pi_{b}\hat{\bf 
b}$ with $\hat{\bf a}$ and $\hat{\bf b}$ being unit vectors along the 
$a$ and $b$ directions, 
\begin{equation}
{\bf v}_{F}\cdot{\bf \Pi}=(v_{Fx}+cv_{Fy})\Pi_{x}
+(v_{Fy}+cv_{Fx})\Pi_{y}\;,
\end{equation}
with $v_{Fx,y}=k_{Fx,y}/m$ in the $x$-$y$ coordinate system. 
The pairing interaction in this coordinate system can be written as
\begin{equation}
V(\phi,\phi^{\prime})=V_{s}+V_{d}\sin(2\phi)\sin(2\phi^{\prime}),
\label{pairing}
\end{equation}
where $\phi=\tan^{-1}(k_{y}/k_{x})$.

For the singlet pairing, $f=f_{0} i\sigma_{y}$ and 
$\Delta=\Delta_{0} i\sigma_{y}$. 
Using the inverse 
of the operator $\hat{L}_{op}=\hat{L}+i \mbox{sgn}(\omega)\mu_{0}^{*}H$, 
which admits the representation
\begin{equation}
\hat{L}_{op}^{-1}=\int_{0}^{\infty} ds \exp (-s\hat{L}_{op}) \;, 
\end{equation} 
we show 
\begin{equation}
f_{0}=2\pi \int_{0}^{\infty} ds 
e^{-s[\hat{L}+2i\mbox{sgn}(\omega)\mu_{0}^{*}H]}\Delta_{0}\;.
\label{anomalous}
\end{equation}
Substituting Eq.~(\ref{anomalous}) into Eq.~(\ref{gap}),    
we obtain the linearized gap equation of an anisotropic mixed $d$- and 
$s$-wave superconductor as follows:
\begin{eqnarray}
\Delta_{0}({\bf R},\phi)=&&2\pi T\sum_{\omega}\langle
V(\phi,\phi')\int{}d\xi e^{-\xi[2|\omega|
+i\mbox{\small sgn}\omega{\bf v}_{F}\cdot{\bf \Pi}]}
\nonumber\\
&&\times\cos(2\mu^{*}_{0}H\xi)\Delta_{0}({\bf R},\phi')\rangle_{SF}.
\end{eqnarray}
It is easy to see that $\Delta_{0}({\bf R},\phi)$ turns out to be
$\Delta_{s}({\bf R})+\Delta_{d}({\bf R})\sin(2\phi)$ because of the 
pairing interaction $V(\phi,\phi^{\prime})$ in Eq.~(\ref{pairing}).

Let us, first of all, consider equations to determine $T_{c}$ of such 
a superconductor. Setting $H=0$, we obtain
\begin{equation}
\Delta_{s}=N_{0}V_{s}\left( 1+{c^2\over2}\right)
\ln{2e^{\gamma}\omega_{D}\over{\pi T_{c}}}\Delta_{s}-
N_{0}V_{s}\left( {c\over2}\right)
\ln{2e^{\gamma}\omega_{D}\over{\pi T_{c}}}\Delta_{d},
\end{equation}
and
\begin{equation}
\Delta_{d}=-N_{0}V_{d}\left( {c\over2}\right)
\ln{2e^{\gamma}\omega_{D}\over{\pi T_{c}}}\Delta_{s}
+N_{0}V_{d}\left({1\over2}+{3\over8}c^{2}\right)
\ln{2e^{\gamma}\omega_{D}\over{\pi T_{c}}}\Delta_{d}.
\end{equation}
In order to calculate $T_{c}$ of an anisotropic
mixed $d$- and $s$-wave superconductor, it is convenient 
to introduce the transition temperature $T_{d(s)}$ 
of the anisotropic $d(s)$-wave
superconductor such that
$N_{0}V_{s}\left( 1+{c^2\over2}\right)
\ln{2e^{\gamma}\omega_{D}\over{\pi T_{s}}}=1$,
and
$N_{0}V_{d}\left( {1\over2}+{3\over8}c^{2}\right)
\ln{2e^{\gamma}\omega_{D}\over{\pi T_{d}}}=1$.
Then, $T_{c}$ can be expressed in terms of $T_{d(s)}$ as follows:
\begin{equation}
\left( 1+{c^2\over2}\right)\left( 
{1\over2}+{3\over8}c^{2}\right)\ln{T_{s}\over 
T_{c}} \ln{T_{d}\over T_{c}}=\left({{\tilde c}\over2}\right)^{2},
\end{equation}
where ${\tilde c}=c\sum_{n=0}^{n_{D}}(n+1/2)^{-1}$ with
$\omega_{D}=(2n_{D}+1)\pi T_{c}$.
Here we would like to point out that a phenomenological value 
$T_{c}\simeq {1\over2}(T_{d}+T_{s})+
{1\over2}\sqrt{(T_{d}-T_{s})^{2}+8{\tilde c}^{2}T_{d}T_{s}}$
can be achieved only if $T_{s}\simeq T_{d}\simeq T_{c}$.
If $T_{s}<<T_{d}$, $T_{c}$ has to be numerically calculated.

Following Ref.~\cite{kim}, we expand
$\Delta_{d(s)}({\bf R})$ in terms of $\{f_{N}({\bf R})\}$,
namely, $\Delta_{d,s}({\bf R})=\sum_{N}A_{N}^{(d,s)}f_{N}({\bf R})$,  
to obtain
\begin{eqnarray}
\Delta_{s}({\bf R})&&=V_{s}\sum_{N}A_{N}^{(s)}
\int{}d\xi
\sum_{m}\sum_{n=0}^{N}\Phi_{n,m}(\xi,N)C_{n,m}^{(s)}(\xi,c)
f_{N-n+m}({\bf R})
\nonumber\\
&&+V_{s}\sum_{N}A_{N}^{(d)}
\int{} d\xi
\sum_{m}\sum_{n=0}^{N}\Phi_{n,m}(\xi,N)C_{n,m}^{(sd)}(\xi,c)
f_{N-n+m}({\bf R})\;,
\label{gap-s}
\end{eqnarray}
and
\begin{eqnarray}
\Delta_{d}({\bf R})&&=V_{d}\sum_{N}A_{N}^{(s)}
\int{}d\xi
\sum_{m}\sum_{n=0}^{N}\Phi_{n,m}(\xi,N)C_{n,m}^{(sd)}(\xi,c)
f_{N-n+m}({\bf R})
\nonumber\\
&&+V_{d}\sum_{N}A_{N}^{(d)}
\int{}d\xi
\sum_{m}\sum_{n=0}^{N}\Phi_{n,m}(\xi,N)C_{n,m}^{(d)}(\xi,c)
f_{N-n+m}({\bf R})\;,
\label{gap-d}
\end{eqnarray}
where 
\begin{eqnarray}
\Phi_{n,m}(\xi,N)=&&2\pi TN_{0}\sum_{\omega}e^{-2|\omega|\xi}
e^{-\vert e\vert H\left( 
{v_{F}\xi\over\sqrt{2}}\right)^2}\cos(2\mu^{*}_{0}H\xi) \nonumber\\
&&\times{1\over\sqrt{m!n!}}
\begin{small}
\left(\begin{array}{c}
             N \\n
\end{array}\right)^{1/2}
\left(\begin{array}{c}
             N-n+m\\m
\end{array}\right)^{1/2}
\end{small}
\bigl[-\vert e\vert H(v_{F}\xi)^{2}\bigr]^{(n+m)/2}
\nonumber
\end{eqnarray}
if $(n+m)$ is even and $\Phi_{n,m}(\xi,N)=0$ if $(n+m)$ is odd, 
and
\begin{eqnarray}
C_{n,m}^{(j)}(\xi,c)=&&
\int{}{d\phi\over 2\pi}{\sin(2\phi)^{j}\over[1+c\sin(2\phi)]}
\exp\left\{ -\vert e\vert H\left({v_{F}\xi\over\sqrt{2}}\right)^{2}
[2c\sin(2\phi)+c^2]\right\}
\nonumber\\
&&\times \bigl(e^{-i\phi}-ice^{i\phi}\bigr)^{m}
\bigl(e^{i\phi}+ice^{-i\phi}\bigr)^{n}
\nonumber
\end{eqnarray}
with $C_{n,m}^{(0)}\equiv C_{n,m}^{(s)}$, 
$C_{n,m}^{(1)}\equiv C_{n,m}^{(sd)}$
and $C_{n,m}^{(2)}\equiv C_{n,m}^{(d)}$.
%%%%%%%%%%%%%%%%%%%%%%
(See Appendix A for the detailed derivation.) 
%%%%%%%%%%%%%%%%%%%%%%
It is necessary to investigate the symmetry properties
of $C_{n,m}^{(j)}$ to calculate $H_{c2}$ as we have mentioned early.
Since $c$ is a small quantity, we may expand the integrand of $C_{n,m}^{(j)}$
up to the $c^{2}$ order.
After a careful investigation of $C_{n,m}^{(j)}$, we find 
\begin{equation}
C_{n,m}^{(0)}\propto {\sin[(m-n)\pi]
\over{(n-m+4)(n-m+2)(n-m)(n-m-2)(n-m-4)}},
\end{equation}
\begin{equation}
C_{n,m}^{(1)}\propto {C_{n,m}^{(0)}
\over{(n-m+6)(n-m-6)}},
\end{equation}
and
\begin{equation}
C_{n,m}^{(2)}\propto {C_{n,m}^{(1)}
\over{(n-m+8)(n-m-8)}}.
\end{equation}
As one can easily see, $C_{n,m}^{(0)}\ne 0$ only if $|n-m|=0, 2, 4$,
$C_{n,m}^{(1)}\ne 0$ only if $|n-m|=0, 2, 4, 6$,
and $C_{n,m}^{(2)}\ne 0$ only if $|n-m|=0, 2, 4, 6, 8$.
We also find other symmetry
properties of $C_{n,m}^{(j)}$ such as $C_{n,m}^{(j)}=-C_{m,n}^{(j)}$
if $|n-m|=2, 6$, and $C_{n,m}^{(j)}=C_{m,n}^{(j)}$ if $|n-m|=4, 8$. 
These properties are still valid even if we expand $C_{n,m}^{(j)}$
up to the $c^{4}$ order.
We, thus, know that $|2N\rangle_{d(s)}$ 
do not couple to $|2N+1\rangle_{d(s)}$ but to $|2N\rangle_{d(s)}$,
and vice versa. Consequently, we expect
that the order parameter is represented by
$\sum_{N}[A_{2N}^{(d)}|2N\rangle_{d}\sin(2\phi)
+A_{2N}^{(s)}|2N\rangle_{s}]$ 
near $H_{c2}$;
in other words, $H_{c2}$ is determined by such coefficients as 
$A_{2N}^{(d)}$ and $A_{2N}^{(s)}$.

Using the orthonormality of $\{f_{N}\}$, we obtain the equation for the
coefficients $A_{N}^{(s)}$ and $A_{N}^{(d)}$ from which $H_{c2}$ will be
calculated as follows:
\begin{eqnarray}
A_{m}^{(s)}&&=V_{s}\sum_{N}A_{N}^{(s)}\int{}d\xi\sum_{n=0}^{N}
\Phi_{n,m+n-N}(\xi,N)C_{n,m+n-N}^{(0)}
\nonumber\\
&&+V_{s}\sum_{N}A_{N}^{(d)}\int{}d\xi\sum_{n=0}^{N}
\Phi_{n,m+n-N}(\xi,N)C_{n,m+n-N}^{(1)},
\end{eqnarray}
and
\begin{eqnarray}
A_{m}^{(d)}&&=V_{d}\sum_{N}A_{N}^{(s)}\int{}d\xi\sum_{n=0}^{N}
\Phi_{n,m+n-N}(\xi,N)C_{n,m+n-N}^{(1)}
\nonumber\\
&&+V_{d}\sum_{N}A_{N}^{(d)}\int{}d\xi\sum_{n=0}^{N}
\Phi_{n,m+n-N}(\xi,N)C_{n,m+n-N}^{(2)}
\end{eqnarray}
with $m+n-N\geq0$, and $|m-N|=0, 2, 4$ for $C_{n,m+n-N}^{(0)}$,
$|m-N|=0, 2, 4, 6$ for $C_{n,m+n-N}^{(1)}$, and 
$|m-N|=0, 2, 4, 6, 8$ for $C_{n,m+n-N}^{(2)}$. The equations to determine
$H_{c2}$ are given by $m=0, 2, 4, \cdots$, namely,
\begin{eqnarray}
A_{0}^{(s)}=&&V_{s}\int{}d\xi\Biggl[
A_{0}^{(s)}\Phi^{0}_{0,0}C_{0,0}^{(s)}
+A_{2}^{(s)}\Phi^{2}_{2,0}C_{2,0}^{(s)}
+A_{4}^{(s)}\Phi^{4}_{4,0}C_{4,0}^{(s)}
\nonumber\\
&&+A_{0}^{(d)}\Phi^{0}_{0,0}C_{0,0}^{(sd)}
+A_{2}^{(d)}\Phi^{2}_{2,0}C_{2,0}^{(sd)}
+A_{4}^{(d)}\Phi^{4}_{4,0}C_{4,0}^{(sd)}\cdots\Biggr]
\nonumber\\
A_{2}^{(s)}=&&V_{s}\int{}d\xi\Biggl[
A_{0}^{(s)}\Phi^{0}_{0,2}C_{0,2}^{(s)}
+A_{2}^{(s)}\sum_{i=0}^{2}\Phi^{2}_{i,i}C_{i,i}^{(s)}
+A_{4}^{(s)}\sum_{i=0}^{2}\Phi^{4}_{2+i,i}C_{2+i,i}^{(s)}\cdots
\nonumber\\
&&+A_{0}^{(d)}\Phi^{0}_{0,2}C_{0,2}^{(sd)}
+A_{2}^{(d)}\sum_{i=0}^{2}\Phi^{2}_{i,i}C_{i,i}^{(sd)}
+A_{4}^{(d)}\sum_{i=0}^{2}\Phi^{4}_{2+i,i}C_{2+i,i}^{(sd)}\cdots\Biggr]
\nonumber\\
A_{4}^{(s)}=&&V_{s}\int{}d\xi\Biggl[
A_{0}^{(s)}\Phi^{0}_{0,4}C_{0,4}^{(s)}
+A_{2}^{(s)}\sum_{i=0}^{2}\Phi^{2}_{i,2+i}C_{i,2+i}^{(s)}
+A_{4}^{(s)}\sum_{i=0}^{4}\Phi^{4}_{i,i}C_{i,i}^{(s)}\cdots
\nonumber\\
&&+A_{0}^{(d)}\Phi^{0}_{0,4}C_{0,4}^{(sd)}
+A_{2}^{(d)}\sum_{i=0}^{2}\Phi^{2}_{i,2+i}C_{i,2+i}^{(sd)}
+A_{4}^{(d)}\sum_{i=0}^{4}\Phi^{4}_{i,i}C_{i,i}^{(sd)}\cdots\Biggr],
\nonumber\\
&&\vdots
\nonumber
\end{eqnarray}
and similar equations of $A_{0}^{(d)}, A_{2}^{(d)}, A_{4}^{(d)}, \cdots$.
$H_{c2}$ is the largest value of solutions which satisfy the condition
for a non-trivial solution to exist in these equations of $A_{N}^{(d,s)}$.  
As we have mentioned, $|2N\rangle_{d(s)} (N=0,1,2,\cdots)$ are involved 
in the 
determination of $H_{c2}$; however, it is expectable that the first few
states such as $|0\rangle_{d(s)}, |2\rangle_{d(s)}$ and $|4\rangle_{d(s)}$
are important because in the case of $c=0$, $|0\rangle_{d},
|2\rangle_{s}$ and $4\rangle_{d}$ play the dominant role in determining
$H_{c2}$. Inclusion of more states such as $|6\rangle_{d(s)}$ and
$|8\rangle_{d(s)}$ gives rise to a difference much less than $1\%$. 

As we did in Ref.~\cite{kim}, we introduce dimensionless unit
in the calculation of $H_{c2}$:  
$t=T/T_{c}$ and $h=2|e|H(v_{F}/2\pi T_{c})^{2}$. 
In addition to
the anisotropy parameter $c$, we also define parameters
$\delta=T_{s}/T_{d}$ and 
$\gamma_{z}=(2\pi\mu^{*}_{0}/ev^{2}_{F})T_{c}$,
which is the strength of spin-magnetic field coupling.  
%%%%%%%%%%%%%%%%%%%%%%%%%%%%%%%%%%%%%%%%%%%%%%%%%%%%%%%
For the sake of comparison with experiment measurement, we convert 
the normalized magnetic field into the dimensional one by using 
\begin{equation}
{H(T)\over{-{dH\over{dT}}|_{T_{c}}T_{c}}}
={h(t)\over{-{dh\over{dt}}|_{1}}}.
\end{equation} 
By solving the eigen-equations for $A^{(d,s)}_{N}$,
we plot $H_c2$ in Fig.~1 as a function of temperature for several typical 
cases. %All calculations are made for a two-dimensional system. 
$H_{c2}$ for an anisotropic mixed $d$- and $s$-wave superconductor with 
ZI taken into account is plotted with the solid line.
$H_{c2}$ for an anisotropic 
mixed $d$- and $s$-wave superconductor  
without ZI is represented by the dot-dashed line. 
The involved parameter values are taken to be: 
$\delta=T_{s}/T_{d}=0.06$, $c=0.16$, and $\gamma_{z}=0.15$.
Also plotted are $H_{c2}$ for a two-dimensional $s$-wave 
supercondcutor (dotted line) and a pure 
$d$-wave superconductor (dashed line) without the MA and the 
ZI. Note that, even though the increase of $\delta$ can enhance 
$H_{c2}$,~\cite{kim} for the value of $\delta$ chosen here, 
the upper critical field for 
an isotropic mixed $d$- and $s$-wave order parameter (with $\delta=0.06$)
without ZI is more or less same as that of the pure $d$-wave superconductor 
(dashed line).  As shown in Fig.~1, we find that the mass anisotropy 
enhances $H_{c2}$ while the spin Zeeman coupling  suppresses $H_{c2}$. 
This means that the mass anisotropy supports superconductivity;
in other words, it increases $T_{c}$ (and consequently the gap), 
which can be easily seen from the equation for $T_{c}$.
%%%%%%%%%%%%%%%%%%%%%%%%%%%%%%%%%%%%%%%%%%%%%%%%%%%%%
In addition, one can also see that $H_{c2}$ is saturated at zero 
temperature as a reflection of the thermodynamic 
requirement; namely, on the phase boundary in the $T-H$ plane,
\begin{equation}
-{dH_{c2}\over{dT}}={\delta S\over{\delta M}}=0
\end{equation}
at $T=0$ near $H_{c2}$, where $\delta S$ is the entropy difference
between the normal and superconducting states, and $\delta M$ is
the magnetization near $H_{c2}$.

Recently, the upper critical field 
$H_{c2}$ (parallel to the $c$ axis) of YBCO
with $T_{c}=84.3\;\mbox{K}$ has been measured 
down to about $4K$.~\cite{nakagawa} The corresponding data is shown by 
solid circles in Fig.~1.
%%%%%%%%%%%%%%%%%%%%%%%%%%%%
The slope $dH/dT|_{T_{c}}$ has been experimentally observed to be
$-1.9\;\mbox{T/K}$.~\cite{welp}
In the theoretical calculation, we are mainly concerned about the 
low-temperature data of $H_{c2}$ most
because at the low temperature thermal fluctuation effect is negligible.
As shown in Fig.~1, the experimental data can be fit very well, 
with the above given parameter values, by our calculation  
for an anisotropic mixed $d$- and $s$-wave superconductor including the ZI.  
Here we would like to point out that, since $\delta$ enhances $H_{c2}$ 
more significantly than $c$,~\cite{kim} 
if we take a large value of $\delta$, we have to choose a physically 
unacceptable high value of $\gamma_z$ to fit the low-temperature data. 
Actually, it is believed that $T_c/E_F \sim 0.1$ 
for a high-$T_{c}$ superconductor,~\cite{kresin} which yields 
$\gamma_z=\pi T_{c}/2E_{F}\approx 0.157$ by assuming $\mu^{*}_{0}=\mu_{B}$ 
(Bohr magneton). On the other hand, most of available experiments seem to indicate that 
$\delta$ should be very small. All this facts demonstrate that the chosen 
set of parameter values are physically reasonable.  
Finally,  we would like to mention: (i) Because 
$H_{p}$ for the sample is about $185\;\mbox{T}$,
the experimental data 
in Ref.~\cite{nakagawa} are within paramagnetic
limit~\cite{clong} with the critical ratio 
$H_{p}/T_{c}$ estimated to be  $2.2\;\mbox{T/K}$ 
for a $d$-wave superconductor. 
Therefore, in our consideration, we do not have to include 
the spin-orbit interaction because
it reduces the pair-breaking effect of the 
Zeeman interaction, and consequently, it allows $H_{c2}$ to
be larger than $H_{p}$.~\cite{tinkham} In the theoretical point of view, 
since the strength of the spin-orbit
coupling is proportional to $Z^{2}$, where $Z$ is an atomic number, 
we may neglect its effect in YBCO as long
as no heavy-atomic impurity is taken into account as in this paper. 
However, it may play an important role
in such heavy-fermion superconductors as UBe$_{13}$ and 
UPt$_{3}$.~\cite{sigrist}
(ii) Magnetic and non-magnetic 
impurities are pair breakers so that it is clear the impurities
reduce $H_{c2}$ as well as $T_{c}$.~\cite{maki} However, the impurity 
concentration in the sample prepared in Ref.~\cite{nakagawa}
seems to be negligible,  the corresponding  effect is not considered here. 
(iii) Recently, O'brien {\it et al.}~\cite{obrien} have interpreted
the experiment based on a three dimensional $s$-wave model.~\cite{whh}
However, it is well-known that superconductivity in YBCO 
is of two dimensional nature.~\cite{cons} 
(vi) A small deviation of theoretical results and experimental 
data occurs in high temperature region because  
thermodynamic fluctuations~\cite{tinkham} of vortices is strong when
temperature is high.

In summary, the upper critical field $(H_{c2})$ of a mixed $d$- and 
$s$-wave superconductor
with a mass anisotropy has been microscopically calculated based on the
quasiclassical theory. We found the mass anisotropy supports $H_{c2}$
against Zeeman suppression. $H_{c2}$ becomes saturated at zero temperature
in consistence with a thermodynamic requirement.
The theoretical results are compared well with
recent experimental data of YBa$_{2}$Cu$_{3}$O$_{7-\delta}$.

One of us (W.K.) would like to thank V. Kogan
and S. C. Lee for helpful discussions.
This work is supported by Texas Center for Superconductivity at the
University of Houston and by the Robert A. Welch Foundation.
%%%%%%%%%%%%%%%%%%%%%%%%%%%%%%%%%
%\newpage
\appendix
\section{}
Since $\Delta_{0}({\bf R},\phi)=\Delta_{s}({\bf R})
+\Delta_{d}({\bf R},\phi)$, we have two equations;
namely, one is for $\Delta_{s}({\bf R})$ and the other 
for $\Delta_{d}({\bf R})$ as follows:
\begin{eqnarray}
\Delta_{s}({\bf R})&&=V_{s}\sum_{\omega}
\int{}{d\phi\over{2\pi}}{1\over{[1+c\sin(2\phi)]}}
\int{}d\xi e^{-2\vert \omega\vert \xi}\cos(2\mu^{*}_{0}H\xi)
\Omega(\Pi_{+},\Pi_{-})\Delta_{s}({\bf R})
\nonumber\\
&&+V_{s}\sum_{\omega}\int{}{d\phi\over{2\pi}}{\sin(2\phi)
\over{[1+c\sin(2\phi)]}}
\int{}d\xi e^{-2\vert \omega\vert \xi}\cos(2\mu^{*}_{0}H\xi)
\Omega(\Pi_{+},\Pi_{-})\Delta_{d}({\bf R})
\end{eqnarray}
and
\begin{eqnarray}
\Delta_{d}({\bf R})&&=V_{d}\sum_{\omega}\int{}{d\phi\over{2\pi}}
{\sin(2\phi)\over{[1+c\sin(2\phi)]}}
\int{}d\xi e^{-2\vert \omega\vert \xi}
\cos(2\mu^{*}_{0}H\xi)
\Omega(\Pi_{+},\Pi_{-})\Delta_{s}({\bf R})
\nonumber\\
&&+V_{d}\sum_{\omega}\int{}{d\phi\over{2\pi}}
{\sin(2\phi)^{2}\over{[1+c\sin(2\phi)]}}
\int{}d\xi e^{-2\vert \omega\vert \xi}
\cos(2\mu^{*}_{0}H\xi)
\Omega(\Pi_{+},\Pi_{-})\Delta_{d}({\bf R}),
\end{eqnarray}
where 
\begin{equation}
\Omega(\Pi_{+},\Pi_{-})=2\pi 
TN_{0}\exp\bigl[-i{\mbox{sgn}}(\omega) 
\left({v_{F}\over\sqrt{2}}\right)\xi
\{(e^{-i\phi}-ice^{i\phi})\Pi_{+}+(e^{i\phi}+ice^{-i\phi})\Pi_{-}\}\bigr]
\end{equation}
with $\Pi_{\pm}=(\Pi_{x}\pm i\Pi_{y})/\sqrt{2}$. Note that
\begin{eqnarray}
&&\bigl[-i{\mbox{sgn}}(\omega)\left({v_{F}\over\sqrt{2}}\right)
\xi(e^{-i\phi}-ice^{i\phi})\Pi_{+},
-i{\mbox{sgn}}(\omega)\left( 
{v_{F}\over\sqrt{2}}\right) \xi(e^{i\phi}+ice^{-i\phi})\Pi_{-}\bigr] 
\nonumber\\ =&&2\vert e\vert H\left( 
{v_{F}\xi\over\sqrt{2}}\right)^{2}[1+2c\sin(2\phi)+c^{2}], 
\end{eqnarray}
then we obtain
\begin{eqnarray}
\Delta_{s}({\bf R})&&=V_{s}\sum_{\omega}\int{}{d\phi\over{2\pi}}
{1\over{[1+c\sin(2\phi)]}}
\int{}d\xi e^{-2\vert \omega\vert \xi} \cos(2\mu^{*}_{0}H\xi)
{\tilde\Omega}(\Pi_{+},\Pi_{-})\Delta_{s}({\bf R})
\nonumber\\
&&+V_{s}\sum_{\omega}\int{}{d\phi\over{2\pi}}
{\sin(2\phi)\over{[1+c\sin(2\phi)]}}
\int{}d\xi e^{-2\vert \omega\vert \xi} \cos(2\mu^{*}_{0}H\xi)
{\tilde\Omega}(\Pi_{+},\Pi_{-})\Delta_{d}({\bf R})
\end{eqnarray}
and
\begin{eqnarray}
\Delta_{d}({\bf R})&&=V_{d}\sum_{\omega}\int{}{d\phi\over{2\pi}}
{\sin(2\phi)\over{[1+c\sin(2\phi)]}}
\int{}d\xi e^{-2\vert \omega\vert \xi}\cos(2\mu^{*}_{0}H\xi)
{\tilde\Omega}(\Pi_{+},\Pi_{-})\Delta_{s}({\bf R})
\nonumber\\
&&+V_{d}\sum_{\omega}\int{}{d\phi\over{2\pi}}
{\sin(2\phi)^{2}\over{[1+c\sin(2\phi)]}}
\int{}d\xi e^{-2\vert \omega\vert \xi}\cos(2\mu^{*}_{0}H\xi)
{\tilde\Omega}(\Pi_{+},\Pi_{-})\Delta_{d}({\bf R}),
\end{eqnarray}
where
\begin{eqnarray}
{\tilde\Omega}(\Pi_{+},\Pi_{-})=&&2\pi TN_{0}
\exp\left[ 
-\vert e\vert 
H\left({v_{F}\xi\over\sqrt{2}}\right)^{2}(1+2c\sin(2\phi)+c^{2})\right]
\nonumber\\
&&\times 
e^{-i{\mbox{sgn}}(\omega)\left( {v_{F}\over\sqrt{2}}\right) \xi(e^{-i\phi}
-ice^{i\phi})\Pi_{+}}
e^{-i{\mbox{sgn}}(\omega)\left( {v_{F}\over\sqrt{2}}\right) \xi(e^{i\phi}
+ice^{-i\phi})\Pi_{-}}.
\end{eqnarray}
Expanding $\Delta_{s(d)}({\bf R})$ in terms of $\{f_{N}({\bf R})\}$ 
and noting 
\begin{equation}
\begin{small}
\left(\begin{array}{c}
             \Pi_{+} \\ \Pi_{-}
\end{array}\right)f_{N}=\sqrt{2\vert e\vert H}
\left(\begin{array}{c}
             \sqrt{N+1}\\ \sqrt{N}
\end{array}\right)
\left(\begin{array}{c}
             f_{N+1}\\ f_{N-1}
\end{array}\right),
\end{small}
\end{equation}
we obtain Eqs.~(\ref{gap-s}) and (\ref{gap-d}).
%%%%%%%%%%%%%%%%%%%%%%%%%%%%%%%%%%

\begin{figure}
\begin{center}
\hspace{0in}
\epsfbox{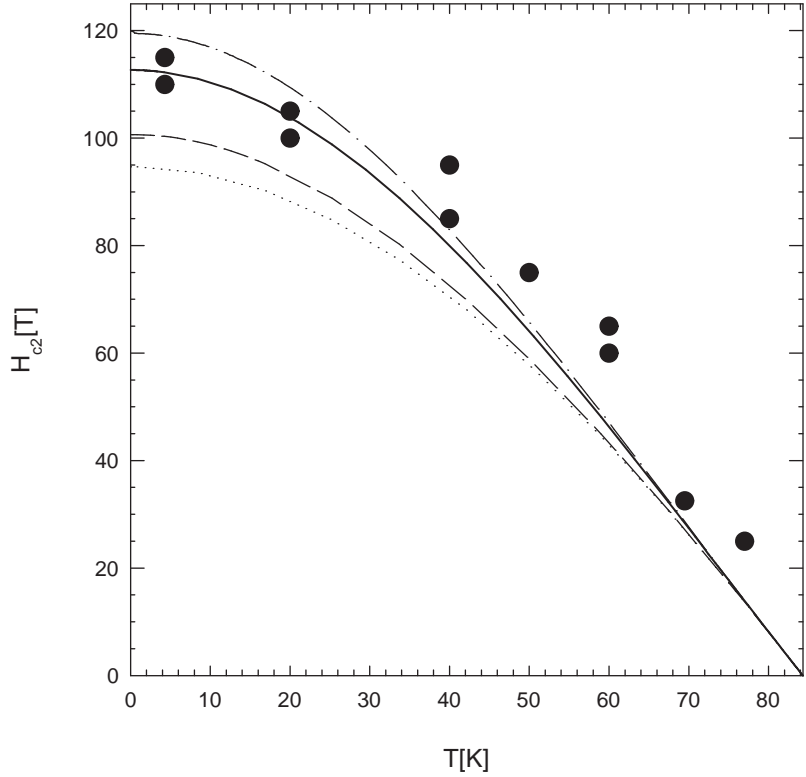}
\end{center}
\caption{The upper critical field $H_{c2}$ for an anisotropic mixed $d$- 
and $s$-wave superconductor (solid line) with the spin Zeeman interaction 
taken into account. Solid circles are experimental data of 
YBCO~\cite{nakagawa}.
Also plotted are $H_{c2}$ of a two-dimensional isotropic $s$-wave 
superconductor (dotted line),
a pure isotropic $d$-wave superconductor (dashed line), 
and an anisotropic  mixed $d$- and $s$-wave superconductor (dot-dashed 
line), all obtained without the Zeeman interaction included.
} 
\end{figure}
\end{document}